\documentclass[9pt]{article}
\usepackage{spconf,amsmath,epsfig}
\usepackage{paralist,amssymb, amsfonts,theorem}
\usepackage{algorithmic}
\usepackage{algorithm}
\usepackage{url}
\usepackage{bm}
\usepackage{amssymb}

\title{PURIFY: a new algorithmic framework for next-generation radio-interferometric imaging}
\name{Rafael E.~Carrillo$^{\star}$, Jason D. McEwen$^{\dagger}$ and Yves Wiaux$^{\ddagger}$}
\address{$^{\star}$Signal Processing Laboratory (LTS5), Ecole Polytechnique F{\'e}d{\'e}rale de Lausanne (EPFL), Switzerland\\
$\dagger$ Mullard Space Science Laboratory, University College London (UCL), UK\\
$\ddagger$ Institute of Sensors, Signals, and Systems, Heriot-Watt University, EH14 4AS, UK}

\begin{document}
%
\maketitle
\begin{abstract}
In recent works, compressed sensing (CS) and convex optimization techniques have been applied to radio-interferometric imaging showing the potential to outperform state-of-the-art imaging algorithms in the field. We review our latest contributions \cite{carrillo12,carrillo13,carrillo13b}, which leverage the versatility of convex optimization to both handle realistic continuous visibilities and offer a highly parallelizable structure paving the way to significant acceleration of the reconstruction and high-dimen-sional data scalability. The new algorithmic structure promoted in a new software PURIFY (beta version) relies on the simultaneous-direction method of multipliers (SDMM). The performance of various sparsity priors is evaluated through simulations in the continuous visibility setting, confirming the superiority of our recent average sparsity approach SARA.
\end{abstract}
\section{Introduction}
\label{sec:intro}
Radio interferometry is a powerful technique that allows observation of the radio emission from the sky with high angular resolution and sensitivity~\cite{thompson01,rau09}. The measurement equation for radio interferometry defines an ill-posed linear inverse problem in the perspective of signal reconstruction. Next-generation radio telescopes, such as the new LOw Frequency ARray (LOFAR), or the future Extended Very Large Array (EVLA)  and Square Kilometer Array (SKA), will achieve much higher dynamic range than current instruments, also at higher angular resolution \cite{rau09}. Also, these telescopes will acquire a massive amount of data, thus posing large-scale problems. Classical imaging techniques developed in the field, such as the CLEAN algorithm and its multi-scale variants \cite{hogbom74,bhatnagar04,cornwell08b}, are known to be slow and to provide suboptimal imaging quality \cite{li11,carrillo12}. This state of things has triggered an intense research to reformulate imaging techniques for radio interferometry in the perspective of next-generation instruments.

The theory of compressed sensing (CS) introduces a signal acquisition and reconstruction framework that goes beyond the traditional Nyquist sampling paradigm~\cite{donoho06,candes06,fornasier11}. Recently, CS and convex optimization techniques have been applied to image deconvolution in radio interferometry~\cite{wiaux09,wiaux09b,wenger10,mcewen11a,li11,carrillo12,hardy13,wolz13} showing promising results. These techniques promise improved image fidelity, flexibility and computation speed over traditional approaches. This speed enhancement is crucial for the scalability of imaging techniques to very high dimensions in the perspective of next-generation telescopes. However, the aforementioned CS-based imaging techniques have only been studied for low dimensional discrete visibility coverages. Therefore, the extension of CS techniques to more realistic continuous interferometric measurements is of great importance.

In this article, we review recent work \cite{carrillo13b} extending the previously proposed imaging approaches in \cite{wiaux09,wiaux09b,carrillo12} to handle continuous visibilities and open the door to large-scale optimization problems. We summarise a general algorithmic framework based on the simultaneous-direction method of multipliers (SDMM) \cite{combettes11} to solve sparse imaging problems. The proposed framework offers a parallel implementation structure that decomposes the original problem into several small simple problems, hence allowing implementation in multicore architectures or in computer clusters, or on graphics processing units. These implementations provide both flexibility in memory requirements and a significant gain in terms of speed, thus enabling scalability to large-scale problems. A beta version of an SDMM-based imaging software written in C and dubbed PURIFY was released that handles various sparsity priors, including our recent average sparsity approach SARA \cite{carrillo12}, thus providing a new powerful framework for radio-interferometric (RI) imaging\footnote{Available at \url{http://basp-group. github.io/purify/}.}. We summarise the performance of different priors through simulations within PURIFY in the continuous visibility setting. Simulation results confirm the superiority of SARA for continuous Fourier measurements. Even though this beta version of PURIFY is not parallelized, we discuss in detail the extraordinary parallel and distributed optimization potential of SDMM, to be exploited in future versions.

\section{Background and Motivation}
\label{sec:BM}
\subsection{State-of-the-art of  CS-based RI imaging algorithms}
\label{ssec:RICS}
CS introduces a signal acquisition framework that goes beyond the traditional Nyquist sampling paradigm~\cite{donoho06,candes06,fornasier11}, demonstrating that sparse signals may be recovered accurately from incomplete data. Consider a complex-valued signal $\bm{x}\in\mathbb{C}^{N}$, assumed to be sparse in some orthonormal basis $\mathsf{\Psi}\in\mathbb{C}^{N\times N}$ with $K\ll N$ nonzero coefficients, and also consider the measurement model $\bm{y}=\mathsf{\Phi}\bm{x}+\bm{n}$, where $\bm{y}\in\mathbb{C}^{M}$ denotes the measurement vector, $\mathsf{\Phi}\in\mathbb{C}^{M\times N}$ is the sensing matrix and $\bm{n}\in\mathbb{C}^{M}$ represents the observation noise. CS provides results for the recovery of $\bm{x}$ from $\bm{y}$ if $\mathsf{\Phi}$ obeys certain properties~\cite{fornasier11}. 

A radio interferometer takes measurements of the radio emissions of the sky, the so-called visibilities. Under restrictive assumptions of narrow-band (i.e. monochromatic) non-polarized imaging on small fields of view, the visibilities measured identify with Fourier measurements \cite{thompson01}. Thus the measurement operator $\mathsf{\Phi}$ essentially reduces to a Fourier matrix sampled on $M$ spatial frequencies. In a realistic continuous visibility setting, one usually has $M>N$ and sometimes $M \gg N$, which will be increasingly the case for next-generation telescopes \cite{rau09}. 

Reconstruction techniques based on CS and convex optimization have been recently proposed for RI imaging. The first application of CS and convex optimization to radio interferometry was reported in \cite{wiaux09}, showing the versatility of the approach and its superiority relative to standard interferometric imaging techniques. After this seminal work others have followed. The works in \cite{wenger10,li11,hardy13} use the following unconstrained synthesis problem to recover $\bm{x}$ from $\bm{y}$:
\begin{equation}\label{bpdn}
\min_{\bar{\bm{\alpha}}\in\mathbb{C}^{N}}\frac{1}{2}\| \bm{y}-\mathsf{\Phi \Psi}\bar{\bm{\alpha}}\|_{2}^{2}+\lambda\|\bar{\bm{\alpha}}\|_{1},
\end{equation}
where $\lambda$ is a regularization parameter that balances the weight between the fidelity term and the $\ell_1$ regularization term. The signal is recovered as $\hat{\bm{x}}=\mathsf{\Psi}\hat{\bm{\alpha}}$, where $\hat{\bm{\alpha}}$ denotes the solution to the above problem. The work in \cite{li11} studied a CS imaging approach based on \eqref{bpdn} and the isotropic undecimated wavelet transform, reporting reconstruction results superior to those of CLEAN and its multi-scale variants. 

As opposed to unconstrained problems such as \eqref{bpdn}, the works in \cite{wiaux09,wiaux09b,mcewen11a,carrillo12,wolz13} proposed to use constrained $\ell_1$ minimization problems of the form
\begin{equation}\label{delta2}
\min_{\bar{\bm{x}}\in\mathbb{R}_{+}^{N}}\|\mathsf{\Psi}^{\dagger}\bar{\bm{x}}\|_{1}
\textnormal{ subject to }\| \bm{y}-\mathsf{\Phi}\bar{\bm{x}}\|_{2}\leq\epsilon,
\end{equation}
where $\mathsf{\Psi}^{\dagger}$ denotes the adjoint operator of $\mathsf{\Psi}$, $\epsilon$ is an upper bound on the $\ell_{2}$ norm of the noise and $\mathbb{R}^{N}_{+}$ denotes the positive orthant in $\mathbb{R}^{N}$, which represents the positivity prior on $\bm{x}$. Unconstrained problems are easier to handle since one of the functions involved in the minimization is differentiable. In fact, there exist fast algorithms to solve such problems, e.g. the FISTA algorithm \cite{beck09}. However, there is no optimal strategy to fix the regularization parameter even if the noise level is known, therefore constrained problems, such as \eqref{delta}, offer a stronger fidelity term when the noise power is known, or can be estimated \emph{a priori}. Hence, we focus our attention on solving problem \eqref{delta} efficiently, especially for very high dimensional problems ($M \gg N$).

\subsection{The SARA algorithm}
\label{ssec:RICS}
Carrillo et al. proposed in \cite{carrillo12} an imaging algorithm dubbed sparsity averaging reweighted analysis (SARA) based on average sparsity over multiple bases, showing superior reconstruction qualities relative to state-of-the-art imaging methods in the field. A sparsity dictionary composed of a concatenation of $q$ bases, 
$\mathsf{\Psi}=[\mathsf{\Psi}_1, \mathsf{\Psi}_2, \ldots, \mathsf{\Psi}_q]$,
with $\mathsf{\Psi}\in\mathbb{C}^{N\times D}$, $N<D$, is used and average sparsity is promoted through the minimization of an analysis $\ell_0$ prior, $\|\mathsf{\Psi}^{\dagger}\bar{\bm{x}}\|_{0} $. The concatenation of the Dirac basis and the first eight orthonormal Daubechies wavelet bases (Db1-Db8) was proposed as an effective and simple candidate for a dictionary in the RI imaging context. See \cite{carrillo13} for further discussions on the average sparsity model, the dictionary selection and other applications to compressive imaging.

SARA adopts a reweighted $\ell_1$ minimization scheme to promote average sparsity through the prior $\|\mathsf{\Psi}^{\dagger}\bar{\bm{x}}\|_{0} $. The algorithm replaces the $\ell_0$ norm by a weighted $\ell_1$ norm and solves a sequence of weighted $\ell_1$ problems where the weights are essentially the inverse of the values of the solution of the previous problem \cite{carrillo12}. The weighted $\ell_1$ problem is defined as:
\begin{equation}\label{delta}
\min_{\bar{\bm{x}}\in\mathbb{R}_{+}^{N}}\|\mathsf{W\Psi}^{\dagger}\bar{\bm{x}}\|_{1}
\textnormal{ subject to }\| \bm{y}-\mathsf{\Phi}\bar{\bm{x}}\|_{2}\leq\epsilon,
\end{equation}
where $\mathsf{W}\in\mathbb{R}^{D\times D}$ denotes the diagonal matrix with positive weights.

\section{A large-scale optimization algorithm}
\label{sec:OP}
In the case of large-scale data problems, i.e. large number of visibilities $M\gg N$, the visibilities may no longer be processed on a single computer but rather in a computer cluster thus requiring a distributed processing of the data for the image reconstruction task. In this distributed scenario we propose to partition the data vector $\bm{y}$ and the measurement operator into $R$ blocks in the following manner:
\begin{equation}\label{blocks}
\bm{y}=
\begin{bmatrix}
\bm{y}_1\\
\vdots \\
\bm{y}_R
\end{bmatrix}
\textnormal{ and  }
\mathsf{\Phi}=
\begin{bmatrix}
\mathsf{\Phi}_1 \\
\vdots \\
\mathsf{\Phi}_R
\end{bmatrix},
\end{equation}
where $\bm{y}_i \in \mathbb{C}^{M_i}$, $\mathsf{\Phi}_i \in \mathbb{C}^{M_i \times N}$ and $M=\sum_{i=1}^R M_i$. Each $\bm{y}_i$ is modelled as $\bm{y}_i=\mathsf{\Phi}_i \bm{x} + \bm{n}_i$, where $\bm{n}_i\in \mathbb{C}^{M_i}$ denotes the noise vector. 

With this partition the optimization problem in \eqref{delta} can be reformulated as
\begin{equation}\label{dbp1}
\min_{\bar{\bm{x}}\in\mathbb{R}_{+}^{N}}\|\mathsf{W\Psi}^{\dagger}\bar{\bm{x}}\|_{1}
\textnormal{ subject to }\| \bm{y}_i-\mathsf{\Phi}_i\bar{\bm{x}}\|_{2}\leq\epsilon_i, i=1,\dots,R,
\end{equation}
where each $\epsilon_i$ is an appropriate bound for the $\ell_2$ norm of the noise term $\bm{n}_i$. In order to solve this nonsmooth problem we need to reformulate it. Note that any convex constrained problem can be formulated as an unconstrained problem by using the indicator function of the convex constraint set as one of the functions in the objective, i.e. $f(\bm{x})=i_{C}(\bm{x})$ where $C$ represents the convex constraint set. The indicator function is defined as $i_{C}(\bm{x})=0$ if $\bm{x}\in C$ or $i_{C}(\bm{x})=+\infty$ otherwise and belongs to the class of convex lower semicontinuous functions. Therefore \eqref{dbp1} can be rewritten as an unconstrained problem of the form
\begin{equation}\label{dbp2}
\min_{\bm{x}\in\mathbb{C}^{N}} f_1(\mathsf{L}_1\bm{x})+\ldots +f_S(\mathsf{L}_S\bm{x}),
\end{equation}
with $S=R+2$. In this formulation $f_1$ and $f_2$ denote the $\ell_1$ sparsity term and the positivity constraint respectively, and $f_3$ to $f_S$ denote the $R$ data fidelity constraints. Thus $\mathsf{L}_1=\mathsf{\Psi}^{\dagger}$, $\mathsf{L}_2=\mathsf{I}$ and $\mathsf{L}_{i+2}=\mathsf{\Phi}_i$ for $i=1,\dots,S$. 

To solve \eqref{dbp2} we use the simultaneous-direction method of multipliers (SDMM), which belongs to the family of proximal splitting methods \cite{combettes11}. Proximal splitting methods proceed by splitting the contribution of each of the functions in \eqref{dbp2} individually so as to yield an easily implementable algorithm. They are called proximal because each non-smooth function is incorporated in the minimization via its proximity operator, which is defined as:
\begin{equation}
\mathrm{prox}_{f}(\bm{x}) \triangleq \arg\min_{\bm{z}\in\mathbb{R}^{N}} f(\bm{z})+\frac{1}{2}\| \bm{x}-\bm{z} \|_2^2,
\end{equation}
where $f$ is convex lower-semicontinous function. Typically, the solution to \eqref{dbp2} is reached iteratively by successive application of the proximity operator associated with each function. SDMM is a generalization of the alternating-direction method of multipliers \cite{boyd10} to a sum of more than two functions. Convergence results of SDMM are based on convergence of the alternating-direction method of multipliers and can be found in \cite{boyd10}.

The SDMM algorithm is summarized in Algorithm~\ref{alg1}. The algorithm is run for a fixed number of iterations, $T_{\rm{max}}$, or until a stopping criteria is met. The algorithm is stopped if the relative variation between the objective function evaluated at successive solutions is smaller than some bound $\xi\in(0,1)$ and if $\| \bm{y}_i-\mathsf{\Phi}_i\hat{\bm{x}}^{(t)}\|_2 \leq \epsilon_i$. The global update (step 5) uses a conjugate gradient algorithm to solve the linear system. Note that steps 7 to 9 in Algorithm \ref{alg1} can be computed in parallel for each $i$. See \cite{carrillo13b} for further details in the derivation of the algorithm and the computation of the proximity operators.
\begin{algorithm}[h!]
\caption{SDMM }\label{alg1}
\begin{algorithmic}[1]
\STATE Initialize $\gamma>0$, $\hat{\bm{x}}^{(0)}$ and $\bm{z}_i^{(0)}=\bm{0}$, $i=1,\dots,S$.
\STATE $\bm{r}_i^{(0)}=\mathsf{L}_i\hat{\bm{x}}^{(0)}$, $i=1,\dots,S$.
\STATE $\bm{x}_i^{(0)}=\mathsf{L}_i^{\dagger}\bm{r}_i^{(0)}$, $i=1,\dots,S$.
\FOR{$t=1,\dots,T_{\rm{max}}$}
\STATE $\hat{\bm{x}}^{(t)}=(\sum_{i=1}^S\mathsf{L}_i^{\dagger}\mathsf{L}_i)^{-1}\sum_{i=1}^S\bm{x}_i^{(t-1)}$.
\FORALL{$i=1,\dots,S$}
\STATE $\bm{r}_i^{(t)}=\mathrm{prox}_{\gamma f_i}( \mathsf{L}_i\hat{\bm{x}}^{(t)}+ \bm{z}_i^{(t-1)})$.
\STATE $\bm{z}_i^{(t)}=\bm{z}_i^{(t-1)} + \mathsf{L}_i\hat{\bm{x}}^{(t)} - \bm{r}_i^{(t)}$.
\STATE $\bm{x}_i^{(t)}=\mathsf{L}_i^{\dagger}(\bm{r}_i^{(t)}-\bm{z}_i^{(t)})$.
\ENDFOR
\IF{$\hat{\bm{x}}^{(t)}$ meets halting criteria}
\STATE Break.
\ENDIF
\ENDFOR
\RETURN $\hat{\bm{x}}^{(t)}$
\end{algorithmic}
\end{algorithm} 

The advantages of this distributed optimization approach are: (i) the visibilities $\bm{y}_i$ and the measurement operators $\mathsf{\Phi}_i$ are local to each node in the cluster, therefore the memory requirements are distributed among  $R$ nodes, with a data dimensionality $M_i \ll M$; (ii) the measurement operators $\mathsf{\Phi}_i$, and their adjoint, are applied locally at each node thus distributing the processing load, for acceleration of the reconstruction process; (iii) the central processing node, where the global update $\hat{\bm{x}}^{(t)}=(\sum_{i=1}^S \mathsf{L}_i^{\dagger}\mathsf{L}_i)^{-1}\sum_{i=1}^S\bm{x}_i^{(t-1)}$ is computed, and the parallel nodes, where the local updates $\bm{x}_i^{(t-1)}$ are computed, only need to exchange information of the size of the image vector at each iteration rather than of the size of the visibilities, thus alleviating the communication requirements to transfer information between nodes. Note that the composite operator $\sum_{i=1}^S\mathsf{L}_i^{\dagger}\mathsf{L}_i=\sum_{i=1}^R\mathsf{\Phi}_i^{\dagger}\mathsf{\Phi}_i + \mathsf{\Psi}\mathsf{\Psi}^{\dagger} + \mathsf{I}=\mathsf{\Phi}^{\dagger}\mathsf{\Phi}+2\mathsf{I}$, needed in the conjugate gradient solver for the global update, can be applied in parallel by each node since $\mathsf{\Phi}^{\dagger}\mathsf{\Phi}=\sum_{i=1}^R\mathsf{\Phi}_i^{\dagger}\mathsf{\Phi}_i$. Although this approach would distribute the processing load of the conjugate gradient step into the parallel nodes, it would incur a communication overhead since each parallel node needs to communicate its result at each iteration of the conjugate gradient algorithm. One approach that can be used to avoid this situation is to precompute and store the composite operator $\mathsf{\Phi}^{\dagger}\mathsf{\Phi}$ in the central processing node. The aforementioned distributed optimization approach could be very appealing for next-generation telescopes where massive amounts of data are acquired. These distributed optimization ideas are not implemented in the beta version of PURIFY and are the subject of ongoing work. The reader is referred to \cite{carrillo13b} for further discussions.

\section{Experimental Results}
\label{sec:Res}
\begin{figure}
    \centering
     
    \includegraphics[trim = 3.4cm 1.1cm 2.1cm 0.5cm, clip, keepaspectratio, width = 4.0cm]{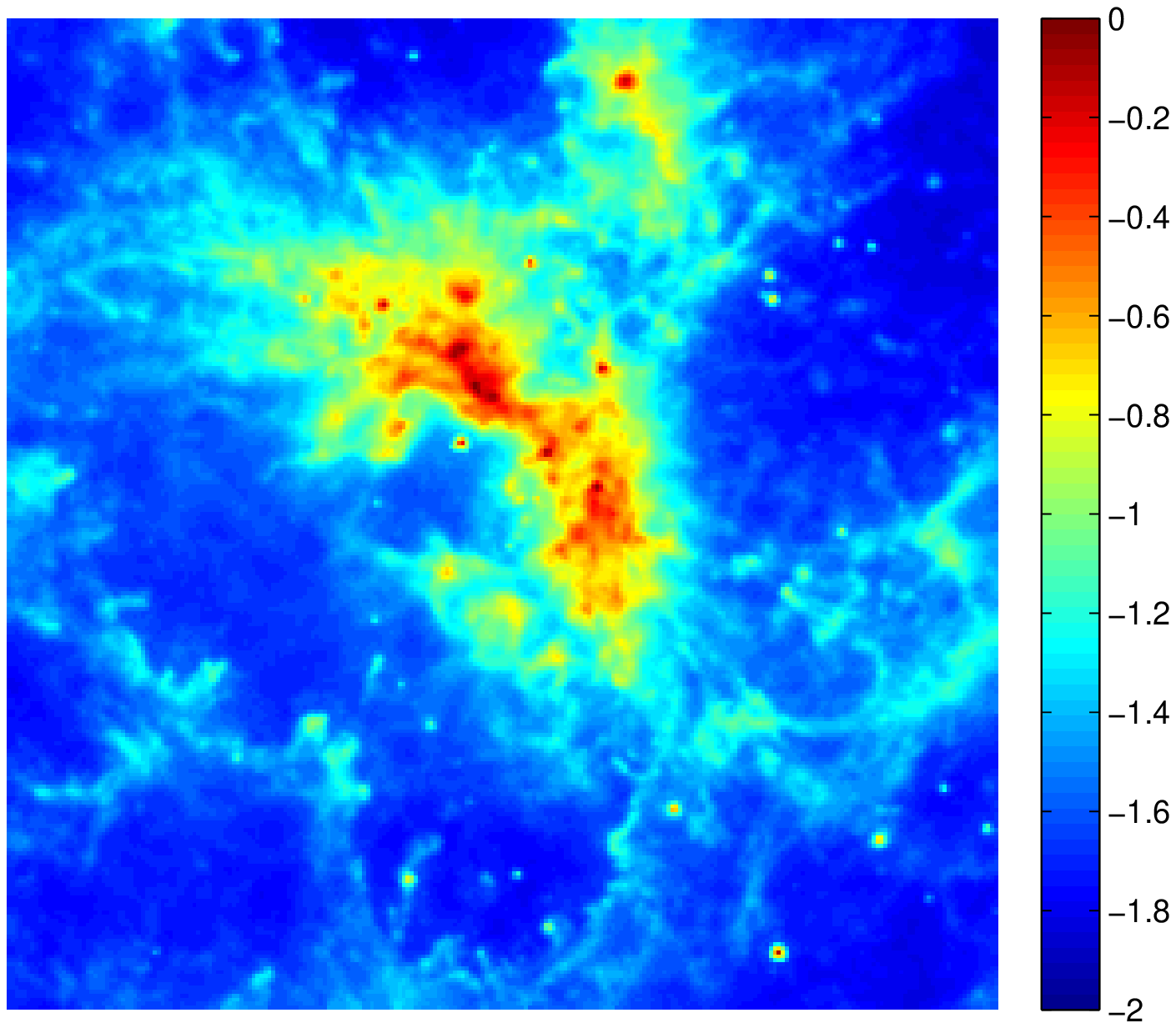}
    \includegraphics[trim = 3.4cm 1.1cm 2.1cm 0.5cm, clip, keepaspectratio, width = 4.0cm]{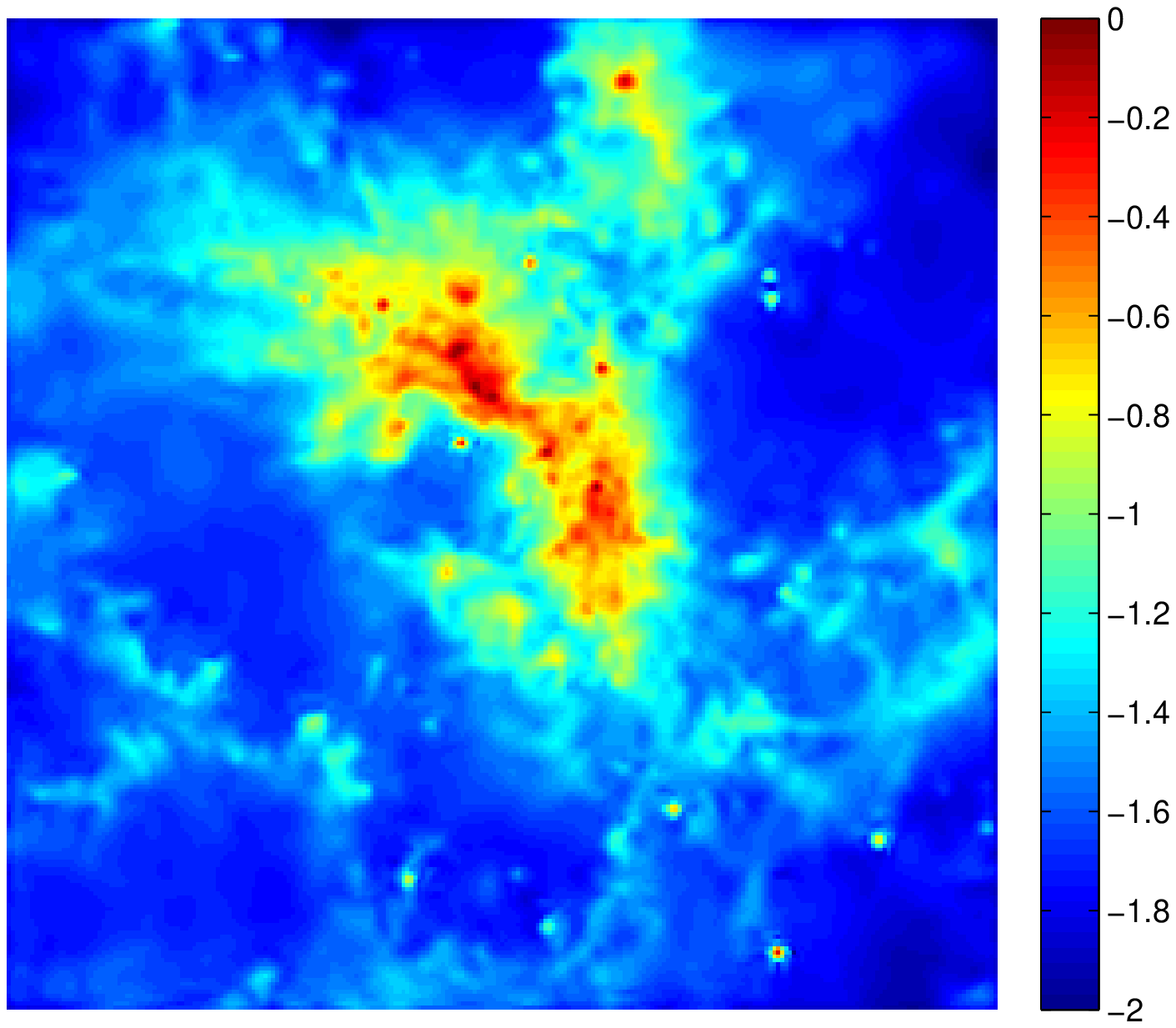}
    
    \includegraphics[trim = 3.4cm 1.1cm 2.1cm 0.5cm, clip, keepaspectratio, width = 4.0cm]{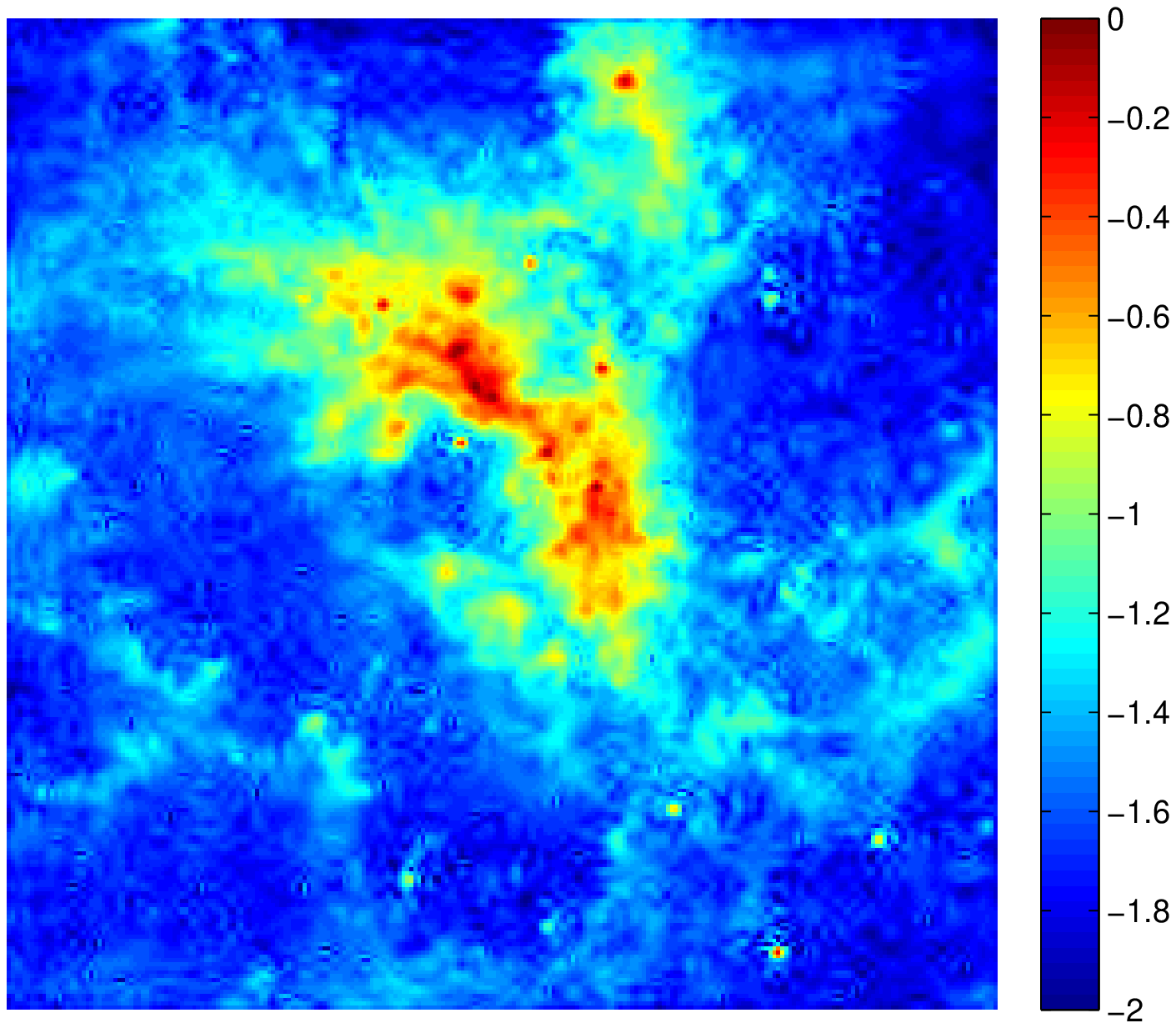}
    \includegraphics[trim = 3.4cm 1.1cm 2.1cm 0.5cm, clip, keepaspectratio, width = 4.0cm]{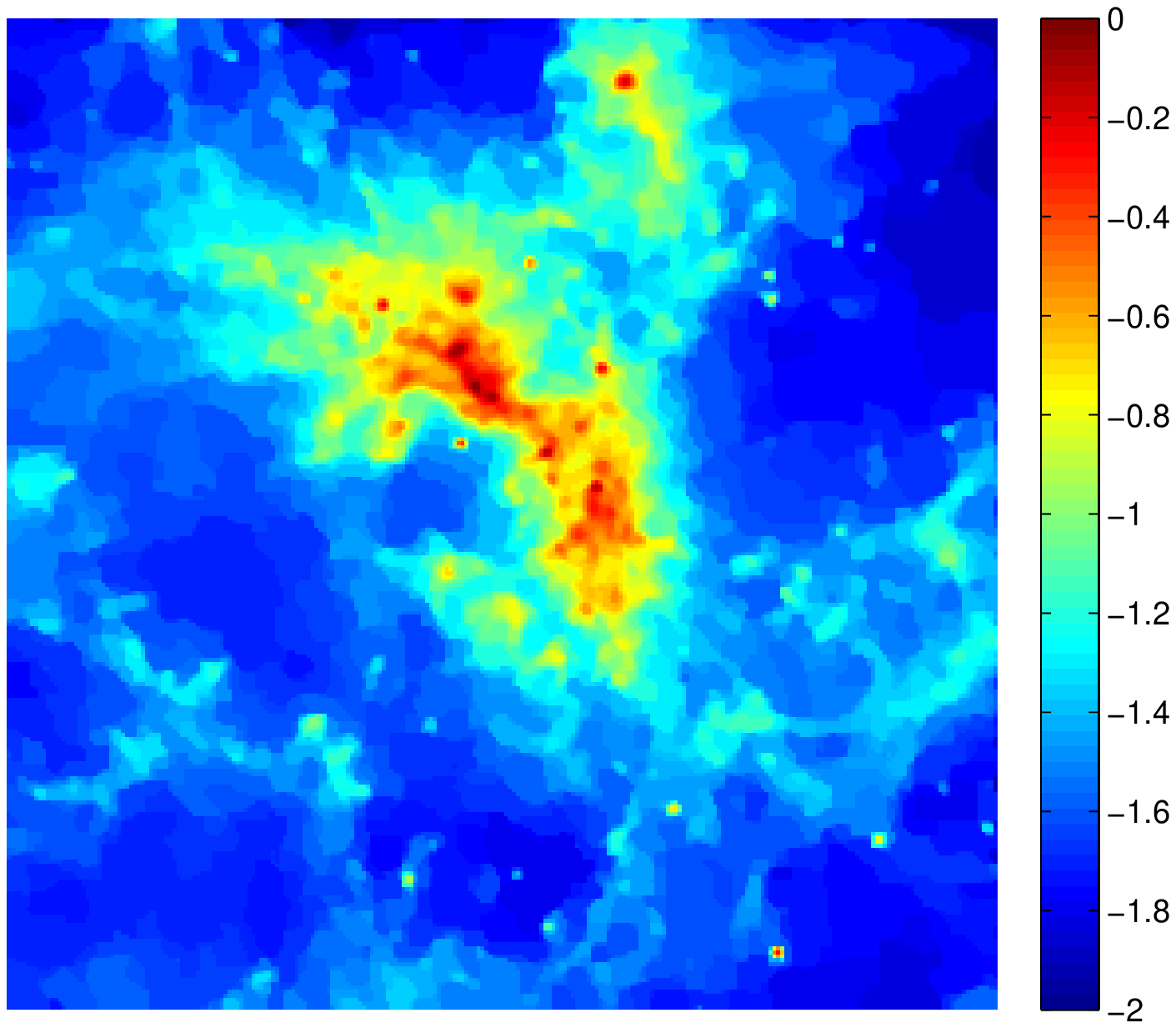}

\caption{Reconstruction example of 30Dor for a coverage with $M=0.4N$ sampling frequencies. Top row left to right: original image and SARA reconstruction (SNR=25.3~dB).
Bottom row left to right: reconstructed images by RWBPDb8 (SNR=22.6~dB) and RWTV (SNR=24.1~dB).}
\label{fig:1}
\end{figure}

In this section we illustrate the performance of the imaging algorithms implemented in PURIFY by recovering the well known 30Dor test image from simulated continuous frequency visibilities. Figure~\ref{fig:1} top-left shows the 256$\times$256  30Dor image used as ground truth image. We use as reconstruction quality metric the signal to noise ratio (SNR). The visibilities are corrupted by complex Gaussian noise with a fixed input SNR (ISNR) set to 30~dB. For the measurement operator, PURIFY implements a non-uniform FFT that maps a discrete image into continuous visibilities \cite{greengard04}. See \cite{carrillo13b} for further details on the measurement operator.

For our evaluation we compare constrained $\ell_1$ and TV minimization problems, as well as their reweighted versions, in terms of reconstruction quality and computation time. For the $\ell_1$ problems we study three different dictionaries $\mathsf{\Psi}$: the Dirac basis, the Daubechies 8 (Db8) wavelet basis and the Dirac-Db1-Db8 concatenation for the SARA algorithm \cite{carrillo12}. The associated algorithms are respectively denoted BP, BPDb8 and BPSA for the non-reweighted case. The reweighted versions are respectively denoted RWBP, RWBPDb8 and SARA. We also study the TV minimization problem with the additional constraint that $\bar{\bm{x}}\in\mathbb{R}_{+}^N$, denoted as TV, and its reweigh-ted version, denoted as RWTV. 

\begin{figure}

\centering
    
    \includegraphics[trim = 1.0cm 0.6cm 0.8cm 0.8cm, clip, keepaspectratio, width = 6.5cm]{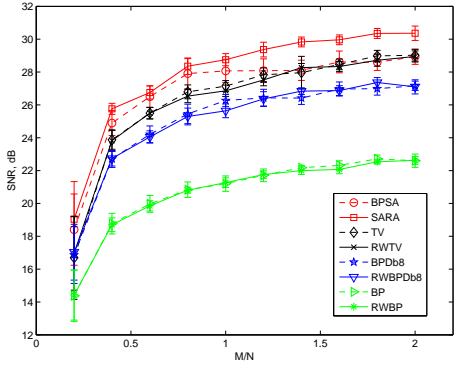}

\caption{Average reconstruction SNR against normalized number of visibilities $M/N$. ISNR is set to 30 dB.}
\label{fig:2}
\end{figure}

In this experiment we use incomplete visibility coverages generated by random variable density sampling profiles. Such profiles are characterized by denser sampling at low spatial frequencies than at high frequencies. This choice mimics common generic sampling patterns in radio interferometry. In order to make the simulated coverages more realistic we suppress the $(0,0)$ component of the Fourier plane from the measured visibilities. This generic profile approach allows us to make a thorough study of the reconstruction quality of the imaging algorithms with a large numbers of simulations for arbitrary number of visibilities and without concern for various telescope configurations. We vary the number of visibilities from $M=0.2N$ to $M=2N$. Reconstruction results for 30Dor are reported in Figure~\ref{fig:2}. Average values over 30 simulations and associated one standard deviation error bars are reported for all plots. The results show that SARA outperforms all other methods in reconstruction quality for the test image. This confirms previous results reported by \cite{carrillo12} in the discrete case now for the more realistic continuous Fourier setting, including the case when $M>N$.

Next we present a visual assessment of the reconstruction quality of the different algorithms. Figure~\ref{fig:1} shows the results for a coverage of $M=26374\approx 0.4N$ visibilities. The reconstructed images are shown in a $\log_{10}$ scale. These images confirm the previous results found by examining recovered SNR levels; SARA yields reconstructed images with fewer artifacts than the other methods. 


\section{Conclusions}
\label{sec:conc}
In this paper we have reviewed a new algorithmic framework based on the simultaneous-direction method of multipliers to solve sparse imaging problems in RI imaging. The new algorithm provides a parallel implementation structure, therefore offering an attractive framework to handle continuous visibilities and associated high dimensional problems. A variety of state-of-the-art sparsity regularization priors, including our recent average sparsity approach SARA, as well as discrete and continuous measurement operators are available in the new PURIFY software. Source code for PURIFY is publicly available. Experimental results confirm both the superiority of SARA for continuous Fourier measurements and the fact that the new algorithmic structure offers a promising path to handle large-scale problems. In future work we will extend the current PURIFY implementation to take full advantage of the parallel and distributed structure of SDMM. Also, direction dependent effects will be included in PURIFY as additional convolution kernels in the operator $\mathsf{\Phi}$ as proposed in \cite{bhatnagar08}. See \cite{wolz13} for first steps in this direction.

\newpage
\bibliographystyle{IEEEbib}
\bibliography{abrev,sara}

\begin{thebibliography}{10}

\bibitem{carrillo12}
R.~E. Carrillo, J.~D. Mc\textsc{E}wen, and Y.~Wiaux,
\newblock ``Sparsity averaging reweighted analysis (\textsc{SARA}): a novel
  algorithm for radio-interferometric imaging,''
\newblock {\em MNRAS}, vol. 426, no. 2, pp. 1223--1234, 2012.

\bibitem{carrillo13}
R.~E. Carrillo, J.~D. Mc\textsc{E}wen, D.~Van~De Ville, J.-P. Thiran, and
  Y.~Wiaux,
\newblock ``Sparsity averaging for compressive imaging,''
\newblock {\em IEEE Signal Process. Letters}, vol. 20, no. 6, pp. 591--594,
  2013.

\bibitem{carrillo13b}
R.~E. Carrillo, J.~D. Mc\textsc{E}wen, and Y.~Wiaux,
\newblock ``\textsc{PURIFY}: a new approach to radio-interferometric imaging,''
\newblock Accepted in MNRAS. Preprint available at
  http://arxiv.org/abs/1307.4370, 2014.

\bibitem{thompson01}
A.~R. Thompson, J.~M. Moran, and G.~W. Swenson,
\newblock {\em Interferometry and Synthesis in Radio Astronomy},
\newblock Wiley-Interscience, New York, 2001.

\bibitem{rau09}
U.~Rau, S.~Bhatnagar, M.~A. Voronkov, and T.~J. Cornwell,
\newblock ,''
\newblock {\em Proc. IEEE}, vol. 97, pp. 1472, 2009.

\bibitem{hogbom74}
J.~A. H\"ogbom,
\newblock ``Aperture synthesis with a non-regular distribution of
  interferometer baselines,''
\newblock {\em A\&A}, vol. 15, pp. 417, 1974.

\bibitem{bhatnagar04}
S.~Bhatnagar and T.~J. Cornwell,
\newblock ``Scale sensitive deconvolution of interferometric images \textsc{I}.
  \textsc{A}daptive \textsc{S}cale \textsc{P}ixel (\textsc{A}sp)
  decomposition,''
\newblock {\em A\&A}, vol. 426, pp. 747, 2004.

\bibitem{cornwell08b}
T.~J. Cornwell,
\newblock ``Multi-scale clean deconvolution of radio synthesis images,''
\newblock {\em IEEE J. Sel. Top. Sig. Process.}, vol. 2, no. 5, pp. 793--574,
  Oct. 2008.

\bibitem{li11}
F.~Li, T.~J. Cornwell, and F.~de~\textsc{H}oog,
\newblock ``Application of compressive sampling to radio astronomy \textsc{I}:
  Deconvolution,''
\newblock {\em A\&A}, vol. A31, pp. 528--538, 2011.

\bibitem{donoho06}
D.~L. Donoho,
\newblock ``Compressed sensing,''
\newblock {\em IEEE Trans. Inf. Theory}, vol. 52, no. 4, pp. 1289--1306, Sept.
  2006.

\bibitem{candes06}
E.~J. Cand\`{e}s,
\newblock ``Compressive sampling,''
\newblock in {\em Proceedings, Int. Congress of Mathematics}, Madrid, Spain,
  Aug. 2006.

\bibitem{fornasier11}
M.~Fornasier and H.~Rauhut,
\newblock {\em Handbook of Mathematical Methods in Imaging}, chapter Compressed
  sensing,
\newblock Springer, 2011.

\bibitem{wiaux09}
Y.~Wiaux, L.~Jacques, G.~Puy, A.~M.~M. Scaife, and P.~Vandergheynst,
\newblock ``Compressed sensing imaging techniques for radio interferometry,''
\newblock {\em MNRAS}, vol. 395, no. 3, pp. 1733--1742, 2009.

\bibitem{wiaux09b}
Y.~Wiaux, G.~Puy, Y.~Boursier, and P.~Vandergheynst,
\newblock ``Spread spectrum for imaging techniques in radio interferometry,''
\newblock {\em MNRAS}, vol. 400, no. 2, pp. 1029--1038, 2009.

\bibitem{wenger10}
S.~Wenger, M.~Magnor, Y.~Pihlstr\"{o}sm, S.~Bhatnagar, and U.~Rau,
\newblock ``Sparse\textsc{RI}: A compressed sensing framework for aperture
  synthesis imaging in radio astronomy,''
\newblock {\em Publ. Astron. Soc. Pac.}, vol. 122, no. 897, pp. 1367--1374,
  2010.

\bibitem{mcewen11a}
J.~D. McEwen and Y.~Wiaux,
\newblock ``Compressed sensing for wide-field radio interferometric imaging,''
\newblock {\em MNRAS}, vol. 413, no. 2, pp. 1318--1332, 2011.

\bibitem{hardy13}
S.~J. Hardy,
\newblock ``Direct deconvolution of radio synthesis images using l1
  minimisation,''
\newblock {\em A\&A}, vol. 557, no. A134, 2013.

\bibitem{wolz13}
L.~Wolz, J.~D. Mc\textsc{E}wen, F.~B. Abdalla, R.~E. Carrillo, and Y.~Wiaux,
\newblock ``Revisiting the spread spectrum effect in radio interferometric
  imaging: a sparse variant of the w-projection algorithm,''
\newblock {\em MNRAS}, vol. 463, no. 3, pp. 1993--2003, 2013.

\bibitem{combettes11}
P.~L. Combettes and J.-C. Pesquet,
\newblock {\em Fixed-Point Algorithms for Inverse Problems in Science and
  Engineering}, chapter Proximal splitting methods in signal processing, pp.
  185--212,
\newblock Springer, New York, 2011.

\bibitem{beck09}
A.~Beck and M.~Teboulle,
\newblock ``A fast iterative shrinkage-thresholding algorithm for linear
  inverse problems,''
\newblock {\em SIAM Journal on Imaging Sciences}, vol. 2, no. 1, pp. 183--202,
  2009.

\bibitem{boyd10}
S.~Boyd, N.~Parikh, E.~Chu, B.~Pelato, and J.~Eckstein,
\newblock ``Distributed optimization and statistical learning via the
  alternating direction method of multipliers,''
\newblock {\em Foundations and Trends in Machine Learning}, vol. 3, no. 1, pp.
  1--122, 2010.

\bibitem{greengard04}
L.~Greengard and J.-Y. Lee,
\newblock ``Accelerating the nonuniform fast \textsc{F}ourier transform,''
\newblock {\em SIAM Review}, vol. 46, no. 3, pp. 443--454, 2004.

\bibitem{bhatnagar08}
S.~Bhatnagar, T.~J. Cornwell, K.~Golap, and J.~M. Uson,
\newblock ``Correcting direction-dependent gains in the deconvolution of radio
  interferometric images,''
\newblock {\em A\&A}, vol. 487, pp. 419, 2008.

\end{thebibliography}

\end{document}